\documentclass{article}
\usepackage{graphicx}

\begin{document}

\begin{center}
{\bf \Large Dynamics of magnetic moments of a nanoscopic array}\\[5mm] 
{\em
{\large A.~Kaczanowski, K.~Malarz and K.~Ku{\l}akowski$^*$}\\[3mm]

Department of Applied Computer Science, 
Faculty of Physics and Nuclear Techniques, 
AGH University of Science and Technology\\
al. Mickiewicza 30, PL-30059 Krak\'ow, Poland.\\[3mm]

$^*$E-mail: kulakowski@novell.ftj.agh.edu.pl\\[3mm]

\today
}
\end{center}

\begin{abstract}
Dynamics of nanoscopic arrays of monodomain magnetic elements is simulated
by means of the Pardavi-Horvath algorithm. Experimental hysteresis loop is
reproduced for the arrays of Ni, with the period 100 nm and the mean coercive
field 710 Oe.We investigate the box-counting fractal dimension of a cluster
of elements with given orientation of magnetic moments. No fractal behavior is
found. Also, the damage spreading technique is applied to check the
criticality. We find that the consequences of a local flip of one magnetic
element remainlimited to a finite area. We conclude that the system does not
show a critical behavior.
\end{abstract}

\section{Introduction} 
Nanoscopic magnetic arrays have been proposed recently as devices of
ultrahigh-density information storage \cite{abra}. On the other hand, it was
announced recently \cite{va1,va2} that an array of bistable magnetic wires
can display a complex behavior. Namely, the authors discussed the effect of 
self-organized criticality (commonly abbreviated as SOC) \cite{bak}.
The problem is that a system in the critical state cannot be used to store any
information, because a flip of one magnetic moment can change the magnetic
structure of the whole array. The aim of this work is to check by means of a
computer simulation if the array is indeed in the critical state. The
values of the simulation parameters are selected as close as possible to the
experimental system described in \cite{abra}.

The dynamics of the total magnetization, i.e. its time dependence in the
presence of the oscillating magnetic field, is governed by two agents: the
long-range magnetostaticinteraction between each two elements of the array,
and the switching field(coercive field), which varies from one element to
another. The nonzeroswitching field enables to preserve information in the
array at positive temperatures. The interaction leads to flips of magnetic
moments and theinformation --- otherwise frozen in the array --- is
destroyed, at leastpartially.  

Let us recall the characteristic features of SOC. According to an
illuminating paperby Flyvbjerg \cite{fly}, excitations wandering in a system
can lead it to a self-organized state, i.e. a spontaneously formed state far
from thermalequilibrium. By criticality we mean that in this state,
excitations can be of any order of magnitude. In fact there is no
characteristic scale in a criticalstate: the array is scale-free in the same
way as a cluster of spins ``up'' in a ferromagnet at its Curie temperature.

Basically, we search for excitations which spread over the lattice. The 
technique applied is known as damage spreading \cite{dam}. It is
believed, that if SOC is present in a system, a local variation of its
structure can lead to a global change \cite{kto}. Then, existence of such a
spreading is an argument for the presence of criticality. On the other hand,
one of the arguments of Refs.~\cite{va1,va2} is based on the calculation of
the box-counting fractal dimension \cite{beck} of the cluster of elements
with a given orientation of magnetic moments. That is why we investigate also
this fractal dimension.     

\section{The system and the Pardavi-Horvath algorithm} 
Our simulations concentrate on the nanoscopic array described in
Ref.~\cite{abra}. Monodomain magnetic cylinders of nickel, 57~nm of
diameter and 115~nm length, form a square array with the period $100$~nm. The
magnetization is $M_s = 370$ emu/cm$^3$. The mean switching field of one
element is $H_s = 710$ Oe, with the standard deviation $105$Oe. The
switching field is constant in time for each element. The magnetostatic
interaction is calculated by means of the rectangular prism approximation
formula \cite{hwa}. The only modification introduced to the investigation of
SOC are the periodic boundary conditions, to avoid boundaries which could
alter the results.

The dynamics of the system is simulated by means of the Pardavi-Horvath (PH) 
algorithm \cite{pho}. Initial parameters of this algorithm are an external magnetic field
$H_{ex}$, the magnetic states of each element of the array and
their switching fields $H_s$. At the begining usually $H_{ex}=0$ and the
magnetic states of the elements are random. Then for every element of the
array, the interaction field $H_i$ can be calculated which comes from other
elements. The numbers which represent this interaction  field are preserved in
the computer memory. After that an element is sought which is most wiling to
flip. This element has to fulfil three conditions: \begin{enumerate}
\item Magnetization of this element must have opposite direction to the total
field $H_{tot}$ exerted on this element ($H_{tot}=H_i+H_{ex}$). \item Total
field must have greater value than the switching field of the element
($\mid H_{tot}\mid >H_s$). \item This element has maximal value of difference
between the total field exerted on the element and its switching field.
\end{enumerate}  The selected element is fliped, and all numbers which
represent the interaction fields are corrected by the change of the interaction
field coming from the flipped element.\\ This procedure is repeated until no
one element can be flipped. Then, the external field is changed.
Subsequent changes of this field are of the values which make possible a flip
of at least one element. The whole procedure is repeated. Finally the
hysteresis loop of the whole array is obtained. 

\begin{figure}[ht] \begin{center}
\includegraphics[angle=-90,width=.9\textwidth]{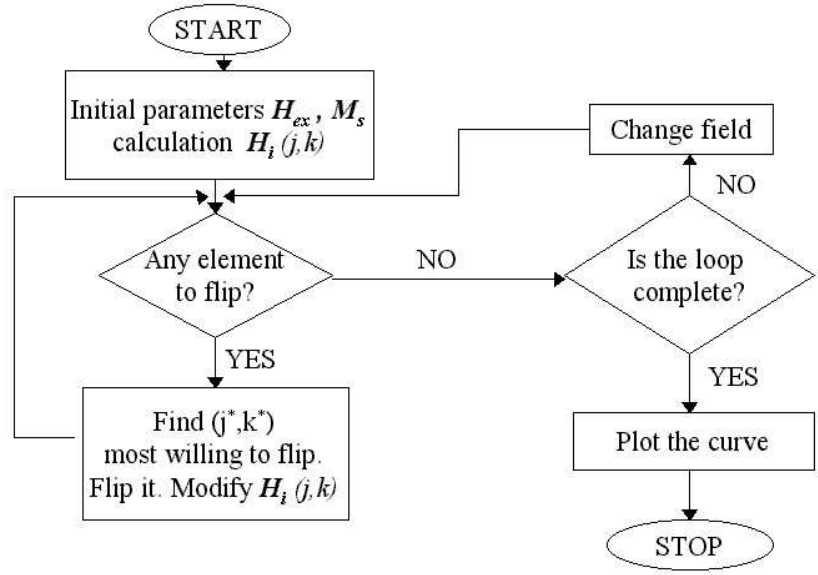} \caption{The
Pardavi-Horvath algorithm \cite{pho}.} \label{P-H}
\end{center} \end{figure}
\begin{figure}[ht]
\begin{center}
\includegraphics[width=.9\textwidth]{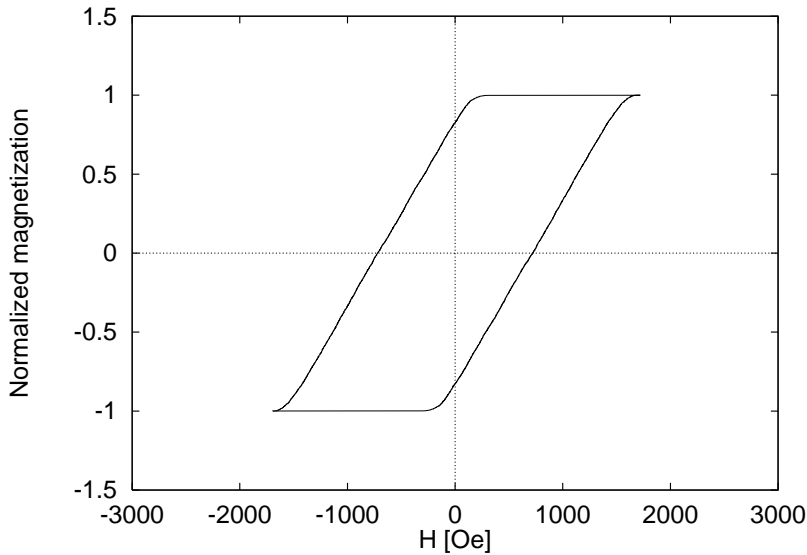}
\caption{The calculated hysteresis loop} \label{FIG1}
\end{center}
\end{figure}

 The algorithm is checked by the calculation of
thehysteresis loop (see Fig.~\ref{FIG1}) and a comparison with the
experimental one \cite{Ross} --- the accordance is quantitative and good. 

\section{Results of simulation} 
 
\subsection{The damage spreading}
\begin{figure}[hbt]
\begin{center}
\includegraphics[angle=-90,width=.9\textwidth]{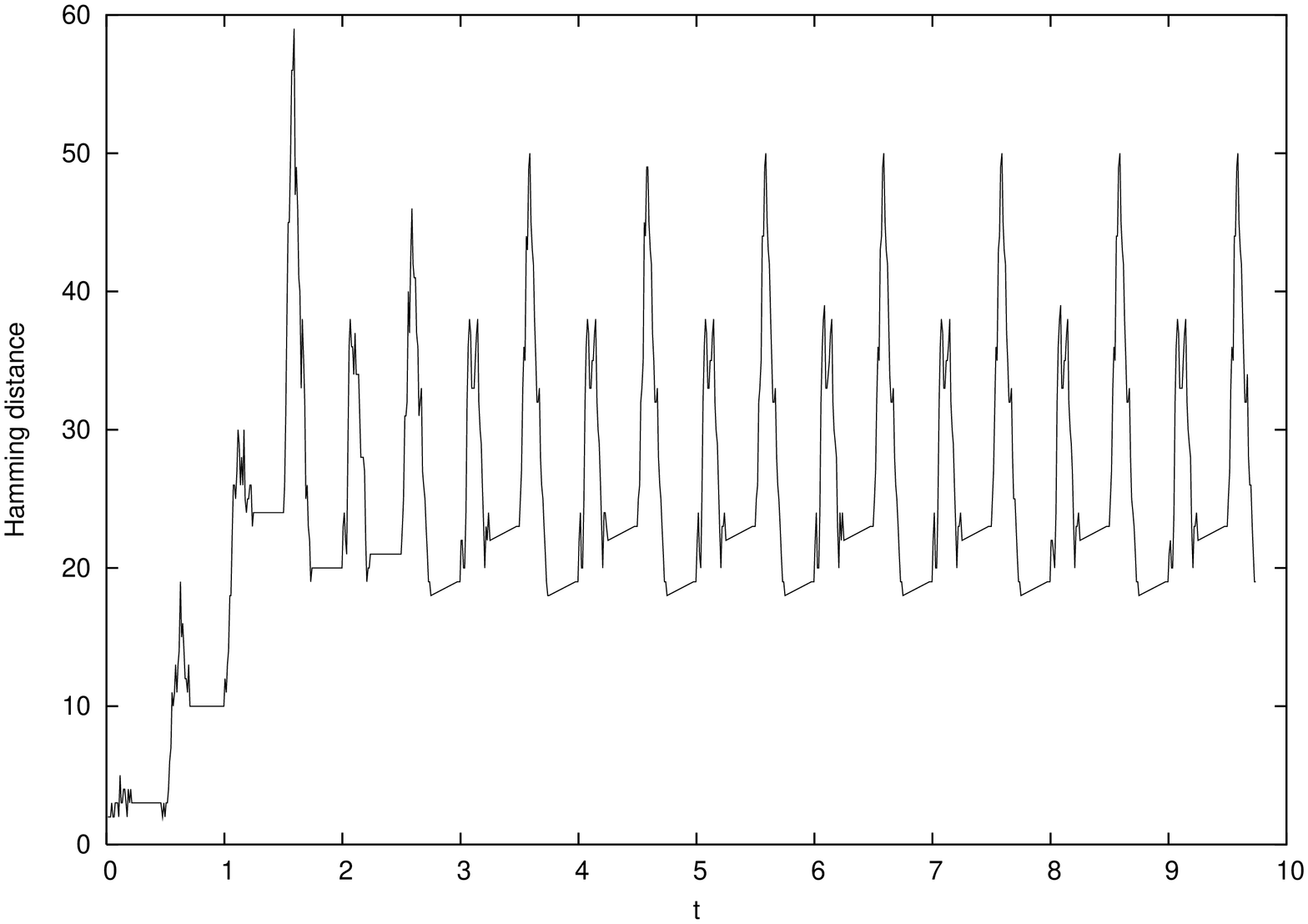}
\caption{Time dependence of the damage.
Maximal value of $A$ in the stationary state is close to fifty.}
\label{FIG2}
\end{center}
\end{figure}

The damage spreading technique is applied for an array of $100\times 100$
elements. Two arrays are stored simultanously in the computer memory, in the
same randomly selected initial state. The PH procedure is applied to lead
these arrays to stable (stationary) states, where no flips occur. Then, one
magnetic moment is flipped in one array, and wecheck if the flip is stable.
Subsequently we apply a periodic magnetic field of amplitude $H_m$. We
consider the case of small frequency, where the whole arrays go to stable
states each time before the field is varied. In this case a particular value
of the frequency is not relevant. The damage is defined as the Hamming
distance between two arrays, defined as the number of elements of the arrays
with opposite directions of the magnetic moments.  

The result is that the Hamming distance increases only during some transient
time. Then, the system reaches a limit cycle, with the length equal to some
multiple of the period of the applied field. A typical result is shown in
Fig.~\ref{FIG2}. We note that if a system is critical, the size of damages
should increase until the system boundaries are reached \cite{kto,kks}. 
In our case damages remain limited to a closed area at the lattice center.

The size $A$ of this maximal damage has a maximum for the amplitude of
the applied field $H_m$ close to 1300 Oe, and it vanishes in most cases for
$H_m > 1400$ Oe. It is obvious that any damage must disappearat strong
fields, when the system is saturated. Besides, $A$ varies strongly from a
sample to asample. As a rule, the damage is localized as a shapeless and
seemingly random pattern of the array elements, formed of several clusters,
separated but close to each other.  

\subsection{The fractal dimension}
\begin{figure}[htbp]
\begin{center}
(a) \includegraphics[angle=-90,width=0.9\textwidth]{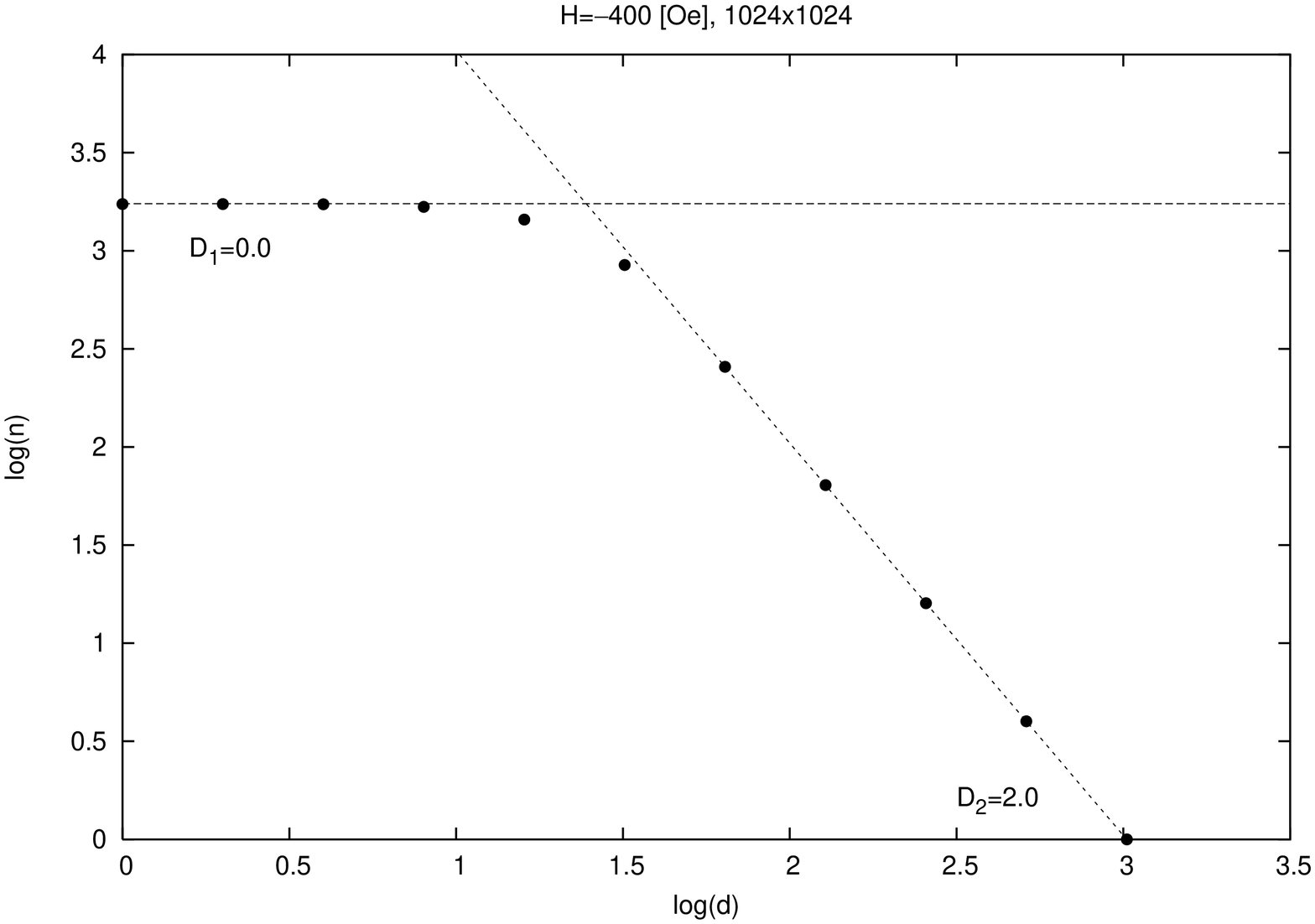}\\
(b) \includegraphics[angle=-90,width=0.7\textwidth]{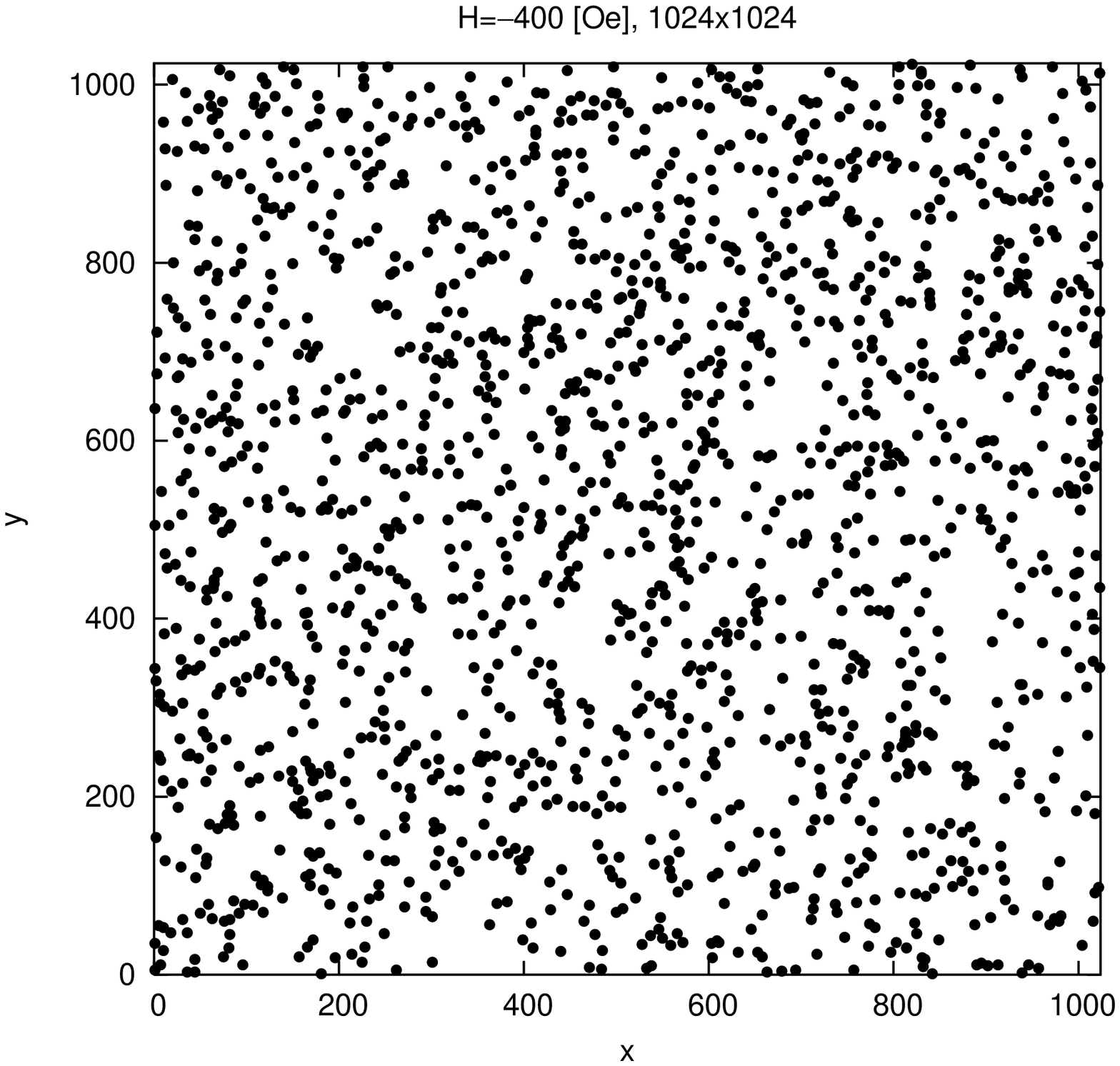}
\caption{
(a) The fractal dimension $D$ as read from the slope $\log(n)$ vs
$\log(d)$, where $n$ is the number of boxes of size $d$, containing
elements with ``up''-oriented magnetic moments.
Here $D$ switches from two to zero.
(b) A typical example of the array configuration for $H=-400$ Oe.
The dots correspond to ``up''-oriented array's elements.}
\label{FIG3}
\end{center}
\end{figure}

The fractal dimension $D$ of the cluster of the ``up'' (or ``down'') oriented
elements is calculated for the $1024\times 1024$ array.Obviously, the
investigated magnetic state of the system should not be arbitrary, but it
should be a consequence of the physical properties and processes. For a
random initial state the fractal dimension $D$ is equal to 2.0, what means
that the system has no fractal character.On the other hand, such a state is an
artifact of the (pseudo)random numbers generator,and therefore it cannot be
treated as realistic.The only physical condition of such a state is its
stability, but in our opinion this condition leaves too much freedom.
Therefore we have calculated the time evolution of the fractal dimension $D$
during the virtually performed hysteresis experiment. It is obvious, that
$D=2.0$ or $D=0.0$ in the saturated states. We found that this bistable
character of $D$ is preserved also for small amplitude $H_m$ of the applied
external field. Between these two values, the scaling is not proper. An
example is shown in Fig.~\ref{FIG3}(a). Typically, a part of the curve for
small size of the box shows the inclination close to zero, and another part
--- close to two. Similar plots have been presented also in
Refs.~\cite{va1,va2}. As the applied field changes, the bent part of the
curve is shifted left and finally disappears.   In fact, plots like
Fig.~\ref{FIG3}(a) appear when we try to assign the  fractal character to a
random configuration of spins. We demonstrate it in Fig.~\ref{FIG3}(b), where
the magnetic configuration of the array produces the curve shown in
Fig.~\ref{FIG3}(a). Surely, this cluster does not exhibit any
self-similarity.  

\section{Conclusions} 
The fractal dimension $D$ obtained in Refs.~\cite{va1,va2} is $1.97$, which is
close to non-fractal value 2.0. On the other hand, in the Ising ferromagnet at
the Curie point $D=187/96\approx 1.95$ \cite{stv}. Then, the value of $D$
itself does not allow to state if the system is in the critical state. 

\begin{figure}[hbt] 
\begin{center} 
\includegraphics[angle=-90,width=.9\textwidth]{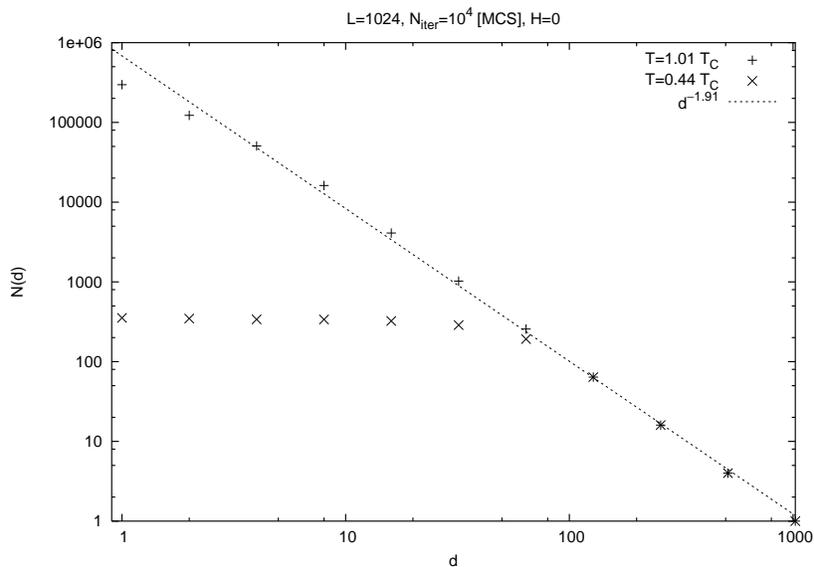} 
\caption{The results of the box-counting technique to the Ising ferromagnet.
Two curves refer to temperatures $1.01T_c$ and $0.44T_c$. The straight line
is a fit for $T=1.01T_c$. The result of the fitting gives $D=1.91$, which is
close to the theoretical value $1.95$ \cite{stv}.}  \label{isi_fd} 
\end{center}  \end{figure}

However, the criterion of criticality can be drawn from the obtained curve.
Its bent character means that there is a characteristic length in the
investigated structure. The value of this length is just at X coordinate of the
bend point. We demonstrate it in Fig.~\ref{isi_fd}. The two curves are obtained
with the same box-counting technique for the conventional Ising ferromagnet.
The curve for temperature close to the $T_c$ is a fairly straight line, except
its upper part where a continuous deviation from linearity is observed. This
deviation can be assigned to a short waiting time (equivalent to $N_{run}$),
which should be very large in the critical region, and to a finiteness of the
system. The curve obtained for $T=0.44 T_c$ displays the same characteristic
length, as in Fig.~\ref{FIG3}(a).This is an indication, that the system is
not scale-free, and therefore not in a critical state.

This conclusion is confirmed by the results obtained by means of the
damage-spreading technique. In all cases investigated, we have not observed a
damage which reaches the boundaries of the system.

However, this conclusion is valid only for the system investigated
experimentally in Ref.~\cite{abra}. We stress this point because there are
serious differences between the this system and the model system discussed in
Refs.~\cite{va1,va2}. In particular, in the latter the lattice is triangular,
what makes the system frustrated because of the antiferromagnetic-like
character of the magnetostatic interaction. Frustration enhances both the
ground state energy and its degeneration. The system can wander in the
phase space, and therefore damages can spread more easily.  Then, the existence
of SOC in an array with the triangular structure remains to be checked. 

\section*{Acknowledgements}
The numerical calculations were partially carried out
in ACK-CYFRONET-AGH.
The machine time is financed by the Ministry of Scientific
Research and Information Technology in Poland, grant No.
KBN/\-HP-K460-XP/\-AGH/\-032/\-2002.

\end{document}